\documentclass[aps,prl,amsmath,amssymb,reprint,floatfix]{revtex4-2}
\usepackage{lmodern}
\usepackage{graphicx}
\usepackage{dcolumn}
\usepackage{bm}
\usepackage[utf8]{inputenc}
\usepackage[T1]{fontenc}
\DeclareUnicodeCharacter{0394}{\ensuremath{\Delta}}
\DeclareUnicodeCharacter{03B1}{\ensuremath{\alpha}}
\DeclareUnicodeCharacter{03B2}{\ensuremath{\beta}}
\DeclareUnicodeCharacter{03BC}{\ensuremath{\mu}}
\DeclareUnicodeCharacter{03C3}{\ensuremath{\sigma}}
\DeclareUnicodeCharacter{03C0}{\ensuremath{\pi}}
\usepackage{mathptmx}
\usepackage{etoolbox}
\usepackage{xcolor}
\usepackage{booktabs}  
\usepackage{multirow}

\usepackage[english]{babel}

\usepackage{hyperref}
\begin{document}

\title{Property-Specific Molecular Representations via Feature-Space Transfer Compression}
\author{Ali Banjafar}
\affiliation{Institut f\"ur Chemie, Universit\"at Kassel, Heinrich-Plett-Stra{\ss}e~40, 34132~Kassel,~Germany}
\author{Guido Falk von Rudorff}%
\email{vonrudorff@uni-kassel.de}
\affiliation{Institut f\"ur Chemie, Universit\"at Kassel, Heinrich-Plett-Stra{\ss}e~40, 34132~Kassel,~Germany}
\affiliation{Center for Interdisciplinary Nanostructure Science and Technology (CINSaT), Heinrich-Plett-Stra{\ss}e 40, 34132 Kassel}

\date{\today}
\begin{abstract}
In many machine learning applications, molecules need to be transformed into representations, i.e. mathematical objects. Those representations are typically considered to be property-agnostic and as such are expected to be over-complete: for different physical properties, different parts or the representation may be relevant. In this work, we propose a method to sub-select and re-weight the representation by adapting it to the property in question. We find that in most cases this makes representations shorter and more accurate at the same time. The feature selection itself uses cheap semi-empirical data instead of high-quality labels. We study four properties (total energy, heat capacity, dipole moment, and polarizability) for three representations (cMBDF, FCHL19, and MACE-MP-0 descriptors) on two datasets (QM9 and VQM24). We can reduce the number of dimensions of a representation in the median by 72\,\% (range 36-98\,\%) while retaining the accuracy. Tuning for accuracy instead we can increase the learning efficiency for dipole moments such that the same accuracy can be reached with 19\,\% of the training data. Our approach yields data-driven interpretations of feature importance, lossless compact representations, and increased data efficiency, requiring only expendable surrogate data.
\end{abstract}

\maketitle
\section{Introduction}

Machine learning models have transformed the field of computational quantum chemistry in recent years~\cite{vonLillienfeld2018, VonLilienfeld2020, Behler2021, Unke2021}. Kernel methods such as kernel ridge regression (KRR) have attracted considerable attention, particularly because they perform very well in low-data regimes~\cite{Montavon2013, Stuke2019, Thant2025}. Kernel-based models require a representation (or descriptor) the role of which is to map atomic information into a vector (global) or a matrix of per-atom descriptors (local)~\cite{Musil2021}. Some representations are built on physical interactions through many-body expansions~\cite{Drautz2019}, such as SOAP~\cite{Bartok2013}, cMBDF~\cite{Khan2023a, Khan2025a}, and FCHL19~\cite{Faber2018, Christensen2020a}; others are derived from pre-trained neural networks~\cite{Behler2007, Schutt2018, Batzner2022, Batatia, Batatia2024} such as MACE-MP\cite{Batatia2025}. These different routes produce long vectors offering a rich bouquet of features to choose from where kernel models consume training data to use the right ones. One alternative to direct learning on this superset of overcomplete features is Automatic Relevance Determination (ARD), which replaces the single kernel length-scale hyperparameter with per-feature length scales~\cite{MacKay1996, Tipping, Rasmussen2008}.

Most molecular representations built so far are designed to universally describe a wide range of molecular properties. Such universal representations contain many features to simultaneously cover these various properties, which naturally results in representations that are over-complete with respect to any single property~\cite{Huang2016, Pozdnyakov2020, Uhrin2021}. This already becomes clear when comparing the formal dimensionality of the molecular Hamiltonian with the representation length: formally, the Hamiltonian has $4N-6+2$ degrees of freedom for $N$ atoms (nuclear charges, spatial coordinates minus translation and rotation, plus spin and net charge), while representations are typically several times longer, sometimes approaching a million features~\cite{Cho2026}. Combining ARD with molecular representations to select features and thereby make the representation property-specific is an open question that many have sought to address~\cite{Fedik2022,Fabregat2022}.

The intrinsic dimensionality of physical properties can be substantially lower even than the formal dimensionality of the Hamiltonian~\cite{Banjafar2025d} which means that many degrees of freedom contribute little to the overall description~\cite{Facco2017}. Motivated by the same intuition --- that removing irrelevant features makes ML models more accurate --- many efforts have been made to construct shorter representations, each successful to some degree. For example, by introducing a square transformation matrix\cite{Fabregat2022} that scales, rotates, and shears the feature space using a fixed kernel length scale and applied to a large, high-quality dataset. Multi-kernel learning\cite{Willatt2018} probes multiple length scales or combines different kernels for separate feature groups. Differentiable Information Imbalance (DII)\cite{Wild2025} and compressive sensing\cite{Nelson2013} has been used to find a subset of features which reproduces distances of a given ground truth. All of these approaches operate at a single level of data quality, on which the model is also evaluated, carrying a risk of overfitting. This distinction is particularly important when the model must operate in low-data regimes, where there are too few data to first select features and then deploy them in the final model.

Transfer learning, which assumes that the correlations and feature relevance in the data carry over across different levels of theory, is by contrast a well-established field in ML for chemistry~\cite{Smith2019, Zaspel2018,Vinod2023,Vinod2025}, typically exploiting the similarity in labels, e.g. via $\Delta$-ML\cite{Ramakrishnan2015} or similarity in features, e.g. for orbital energies\cite{Welborn2018,Karandashev2022,Husch2021}. So far, however, the information contained in the low-quality data has not been used to act on the representation itself --- that is, to weight the features using the low-quality data and then apply the weighted representation to high-quality data.

In this work, we combine transfer learning with ARD kernels to construct property-dependent representations. The kernel feature weights are optimized on either global (vector) and local (matrix) representations to yield a final weight vector that is then applied to the high-quality data.  The weight vector is optimized with the Adam optimizer~\cite{Kingma2017} on the low-quality data~\cite{Bannwarth2019} of two databases, QM9~\cite{Ramakrishnan2014, Ruddigkeit2012} and VQM24~\cite{Khan2024}, treated separately. Throughout the optimization, the high-quality data remain entirely unseen, and the optimizer is fed random mini-batches drawn from the low-quality pool. The optimized weight vector is then applied to the hold-out high-quality data, and its accuracy is compared with the unweighted baseline. Four molecular properties are examined: total energy, heat capacity, dipole moment, and polarizability.

\section{Methods}

\begin{figure}[htbp]
    \centering
    \includegraphics[width=\columnwidth, trim={0.85cm 5.95cm 22.9cm 0.5cm},
    clip]{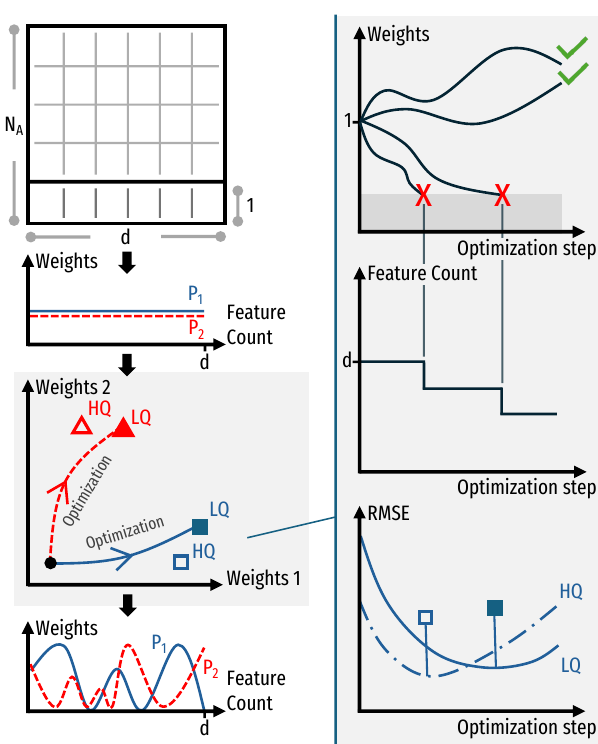}
    \caption{Schematic overview of the method. \textit{Left}: molecular representations are either global (a $d$-dimensional molecular vector) or local (one vector for each of the $N_A$ atoms). The initial weights $\mathbf{w}$ are uniform and then are optimized on low-quality data (LQ). The optimal weights are assumed to be similar for LQ and high quality data (HQ), which enables the transfer. After optimization, weights are property-specific and potentially sparse, yielding compact property-dependent representations. \textit{Right}: Optimization procedure. Feature weights falling below a threshold are eliminated ({\color{red}$\times$}), while informative ones survive ({\color{green}$\checkmark$}); the feature count decreases accordingly. The test RMSE on LQ data decreases monotonically, while the HQ error first improves then rises as overfitting towards LQ appears; the optimal weight vector corresponds to the HQ minimum ($\square$).}
    \label{fig:overview}
\end{figure}

%GUIDO HERE

\subsection{Kernel Ridge Regression}
For a molecule of given nuclear charges $Z_I$ and positions $\mathbf{R}_I$, 
one can find a representation $\mathbf{x}(Z_I, \mathbf{R}_I|\theta) \in \mathbb{R}^d$
for each of $n$ molecules ($\mathbf{X}\in \mathbb{R}^{n\times d}$) governed by some 
representation settings $\theta$. With those representations, we employ 
Kernel-Ridge-Regression (KRR) with a global Gaussian kernel:
\begin{align}
    k^\mathrm{G}(\mathbf{x}, \mathbf{x}') = \exp\left(-\frac{1}{2\sigma^2}\|\mathbf{x}-\mathbf{x}'\|_2^2\right)\label{eqn:kgaussianvanilla}
\end{align}
where $\sigma$ is the length-scale hyperparameter. A KRR model $f(\mathbf{x}_\mathrm{q})$ is 
obtained from $n_\mathrm{t}$ training data points $\mathbf{X}_\mathrm{t} \in \mathbb{R}^{n_\mathrm{t}\times d}$:
\begin{align}
    f(\mathbf{x}_\mathrm{q}) = \sum_{i=1}^{n_\mathrm{t}} \alpha_i\, k^\mathrm{G}(\mathbf{x}_\mathrm{q}, \mathbf{x}_i)
\end{align}
where $\mathbf{x}_\mathrm{q}$ is the representation of the query molecule and the model 
coefficients $\boldsymbol{\alpha}$ are found by minimizing the loss function
\begin{align}
    \mathcal{L} = \sum_{i=1}^{n_\mathrm{t}} (f(\mathbf{x}_i)-y_i)^2 + \lambda \|f\|^2
    \qquad;\qquad \|f\|^2\equiv\boldsymbol{\alpha}^T\mathbf{K}\boldsymbol{\alpha}
\end{align}
with the \textit{RKHS norm} $\|f\|^2$ regularizing the model. The model for a \textit{fixed} 
representation $\mathbf{x}(Z_I, \mathbf{R}_I)$ can be obtained from
\begin{align}
    \boldsymbol{\alpha} = (\mathbf{K}+\lambda \mathbf{I})^{-1}\mathbf{y}
\end{align}
Typically, a fixed universal (i.e. property-independent) representation is assumed for which the length scale $\sigma$ and the regularization strength $\lambda$ are determined through hyperparameter optimization. We now explore whether making the representation property-dependent can improve the prediction accuracy and reduce the feature vector length. Shorter and more relevant feature vectors are expected to yield more data-efficient models.

\subsection{Feature weights}
We introduce per-feature weights $\mathbf{w}\in \mathbb{R}^d$ via the kernel
\begin{align}
    k^\mathrm{G}_\mathbf{w}(\mathbf{x}, \mathbf{x}') = \exp\left(-\sum_{i=1}^d w_i^2(x_i-x'_i)^2
    \right)\label{eqn:gaussianard}
\end{align}
which recovers eqn.~\ref{eqn:kgaussianvanilla} for $w_i^2 = 1/(2\sigma^2)$. The added flexibility is an implied per-feature length scale $\sigma_i = \sqrt{1/(2w_i^2)}$ and the aspect that features of vanishing $\mathbf{w}_i$ are irrelevant and can be dropped without consequence. Crucially, KRR does not fail on irrelevant (or even random) features as long as the relevant features are also present\cite{Shawe-Taylor2005}: the kernel will still encode the correct similarity structure. However, irrelevant features reduce the sensitivity of kernel to the relevant structure in the data effectively reducing the data-efficiency of the model. Dropping features with vanishing weights therefore directly reduces data requirements, improving sample efficiency without sacrificing predictive power.

Our approach operates on two sets of data for the same molecular property over the same chemical space: a large pool of low-quality data $\mathcal{D}_\mathrm{LQ} = \{(\mathbf{x}_i, y_i^\mathrm{LQ})\}_{i=1}^{N_\mathrm{LQ}}$ and a much smaller set of high-quality data $\mathcal{D}_\mathrm{HQ} = \{(\mathbf{x}_i, y_i^\mathrm{HQ})\}_{i=1}^{N_\mathrm{HQ}}$. Both sets share the same molecular representations $\mathbf{x}$ but none of the configurations $(Z_I, \mathbf{R}_I)$. The central assumption illustrated in Fig.~\ref{fig:overview} is that feature relevance is more dependent on the property in question rather than on which quality level this property has been calculated on: the subset of features in $\mathbf{x}$ that carries predictive information for a given property is similar across fidelity levels. Under this assumption, the weights $\mathbf{w}$ optimized by minimizing RMSE of low-quality data on the mini-batches transfer directly to high-quality data, yielding a compressed, property-specific representation that improves KRR accuracy on the expensive target without requiring large amounts of high-quality data.

\subsection{Weight Optimization}

In standard ARD, the weights $\mathbf{w}$ are found by maximizing the log-marginal likelihood (LML) of the training data\cite{Rasmussen2008, MacKay1996}. Although the LML penalizes model complexity, it is inherently biased toward the current set of training points and does not guarantee that the selected features generalize beyond them. We therefore instead adopt an \textit{online transfer learning} scheme: the weights are optimized by gradient descent using Adam optimizer \cite{Kingma2017} over successive mini-batches drawn from a large pool of low-quality surrogate data. Since both levels of theory describe the same molecular properties over the same chemical space, the feature relevance learned from the cheap surrogate is expected to transfer to the expensive high-quality target. This scheme simultaneously pursues three goals: improving prediction accuracy on the high-quality target without adding high-quality training data, compressing the representation by driving irrelevant feature weights toward zero, and yielding property-dependent weights that make the representation itself property-specific.

The weights $\mathbf{w}$ are initialized uniformly and are updated to minimize the RMSE on sequential batches of size $n_\mathrm{b}$
\begin{align}
\mathrm{RMSE}_\mathrm{LQ}(\mathbf{w}) = \sqrt{\frac{1}{n_\mathrm{b}}\sum_{i=1}^{n_\mathrm{b}}
\left(f_\mathbf{w}(\mathbf{x}_i) - y_i^\mathrm{LQ}\right)^2}
\end{align}
Cycling over successive mini-batches from the full surrogate pool prevents the weights from overfitting to any fixed subset of the low-quality data. As training progresses, weights of irrelevant features decay toward zero; those falling below a small threshold $\epsilon$ are set exactly to zero and the corresponding features are dropped, yielding a progressively sparser, property-specific representation.

To monitor progress, the intermediate weight vector is applied to a KRR model on a constant training set of the HQ data with own global hyperparameter optimisation but fixed weights. The weight vector that minimizes the validation error is  then selected as the final model; we report the corresponding test error on the HQ hold-out set. Plotting the test error as a function of the number of retained features yields the compression curve (see below), which quantifies the trade-off between feature dimensionality and predictive accuracy.

Since representation features are not standardized across a data set, it is not obvious that the weights alone should drive discarding features. In eqn.~\ref{eqn:gaussianard}, each feature influences not by the weight alone but by the weight together with the spread of that feature across the dataset. 
Each feature therefore contributes to the kernel the expected value of $w_i^2\,\mathrm{Var}(x_i)$, not $w_i^2$ alone. One might consequently rank and drop features by this product, rather than by $w_i$, so that the absolute scale of a feature also decides whether it is relevant. This would be more closely related to sketching or approximating the existing representation space.

We instead select features by the weights $w_i$ directly, for two reasons. The first and more important one concerns low-variance features. A relevant feature with small $\mathrm{Var}(x_i)$ retains a small product $w_i^2\,\mathrm{Var}(x_i)$, so it is more likely to drop it before giving it the opportunity to be amplified by its weight. Selecting by $w_i$ keeps such features, and we observe that several low-variance features are pulled up during optimization and contribute appreciably to the accuracy gain. The improvement therefore does not stem only from dropping irrelevant features, but also from magnifying relevant ones that are underrepresented in the raw-variance (unweighted) representation. The second reason follows from the same bias: because the unweighted kernel is itself dominated by high-variance features, selecting by the product preferentially retains those same high-variance directions, so the compressed kernel stays close to the unweighted one and improves little. Selecting by $w_i$ instead keeps the features the optimization has promoted, reshaping the kernel away from the raw-variance baseline, which is where the gain originates.

\subsection{Representations}
\label{subsec:representations}
Three molecular representations with different design approaches are considered: cMBDF\cite{Khan2023a, Khan2025a}, FCHL19\cite{Faber2018, Christensen2020a}, and the descriptors from the neural network MACE-MP-0 (medium)~\cite{Batatia, Batatia2024}, short \textit{MACE}. Each encodes the local chemical environment of an atom as a per-atom descriptor $\boldsymbol{\mathbf{x}}_I \in \mathbb{R}^d$; the global molecular representation is obtained by summing (for cMBDF: summation after bagging\cite{Khan2025a}) over all $N_A$ atoms. cMBDF encodes pairwise and many-body interatomic interactions through contracted distribution functions. FCHL19 combines two-body and three-body terms with smooth distance-weighted cutoffs. MACE is a pre-trained equivariant message-passing descriptor that implicitly captures higher-order many-body interactions. The dimensionality of cMBDF and FCHL19 scales with the number of chemical elements in the dataset, ranging from $d \approx 1800$--$2100$ and $d \approx 720$--$2200$ respectively; MACE produces a fixed descriptor of $d = 256$ in our settings.

For the local case, the per-atom descriptors $\mathbf{x}_a$ are used directly without summation, and the molecular kernel is defined over sets of atomic environments. The feature weights enter as a feature scaling applied to each per-atom descriptor, $\tilde{\boldsymbol{\mathbf{x}}}_a = \mathbf{w} \odot \boldsymbol{\mathbf{x}}_a$, before kernel evaluation. The \textit{local kernel} aggregates all atom pairs between two molecules $A, B$ regardless of element:
\begin{align}
k^{L}(A,B) = 
    \sum\limits_{i\in A}\sum\limits_{j\in B} k^\mathrm{G}_\mathbf{w}(\mathbf{x}_i, \mathbf{x}_j')
\end{align}
with normalisation
\begin{align}
k^{\mathrm{local}}(A,B) = \frac{
    k^{L(A,B)}
}{\sqrt{k^{L}(A,A)k^{L}(B,B)}}
\end{align}
The \textit{elemental kernel} refines this by restricting the sum to atom pairs of the same chemical element via the Kronecker delta $\delta_{Z_i,Z_j}$:
\begin{align}
k^{E}(A,B) =
    \sum\limits_{i\in A}\sum\limits_{j\in B} \delta_{Z_i,Z_j}\, k^\mathrm{G}_\mathbf{w}(\mathbf{x}_i, \mathbf{x}_j')
\end{align}
and identical normalisation.

\begin{figure*}
    \centering
    \includegraphics[width=\textwidth]{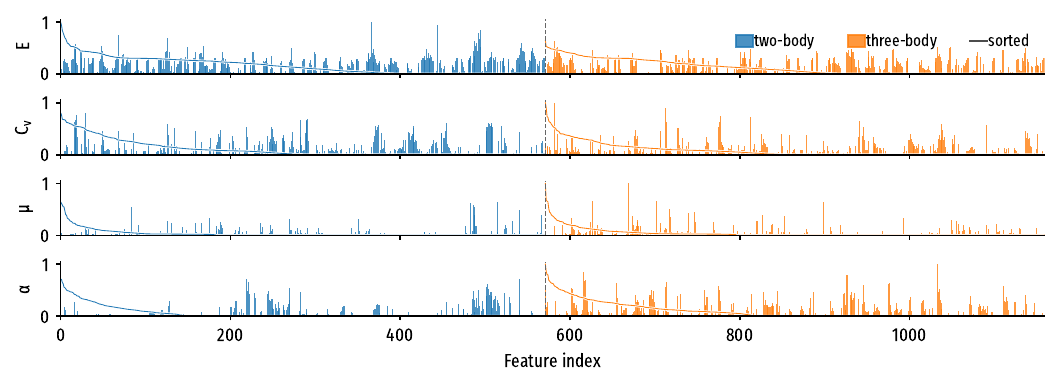}
\caption{Normalized feature weight vectors for cMBDF on QM9, shown for four target properties. Each bar gives the weight of one feature, normalized by the maximum weight for that property. Only those 1160 features assigned nonzero weight by at least one property are shown (571 two-body and 589 three-body out of 1880 total); the dashed line marks the boundary between the two blocks. The overlaid curve shows the sorted weight profile within each block.}
    \label{fig:cMBDF_QM9_weight_heatmap}
\end{figure*}

\subsection{Datasets}
We use two benchmark datasets. QM9\cite{Ramakrishnan2014, Ruddigkeit2012} contains 133,885 small organic molecules containing elements H, C, N, O, and F, with properties computed at B3LYP/6-31G(2df,p) level of theory, of which we consider total energy, heat capacity, dipole moment, and polarizability. VQM24\cite{Khan2024} is a larger and chemically more diverse dataset of 835,947 configurations spanning ten elements (H, C, N, O, F, Si, P, S, Cl, Br), from which total energy and heat capacity are considered. Low-quality labels at the GFN2-xTB level of theory have been pre-computed for QM9\cite{Nandi2023} and we obtained the corresponding labels for VQM24.

For each dataset and property, the individual configurations are strictly partitioned into a high-quality set and a low-quality pool. The low-quality pool is used exclusively for weight optimization; the high-quality set is held out entirely and never seen during optimization, such that neither the labels nor the molecular geometries of the HQ data enter the gradient updates. For HQ data, only the validation error for a constant fixed training set size is used to determine when to stop the weight optimization.  Prior to training, all labels are detrended by subtracting a linear fit on stoichiometry, removing the dominant compositional contribution to the target property.

The same cross-validation procedure\cite{Grimblat2026} is applied in all cases; all improvements are reported as relative reductions in RMSE with respect to the unmodified baseline, so that the method is always compared to the best standard KRR achievable for a given representation.

\subsection{Calculation settings}
All experiments use the Adam optimizer with a learning rate of $\eta = 0.05$. The learning rate controls the step size of each gradient update; values that are too large risk dropping relevant features prematurely by pushing multiple weights below the threshold in a single step, while smaller values slow convergence without affecting the final result. For global representations, the low-quality mini-batch size is $n_\mathrm{b} = 2048$ and the high-quality training set size is $N_\mathrm{HQ} = 1024$. For local representations, a smaller mini-batch of $n_\mathrm{b} = 512$ is used due to the higher computational cost of local kernel evaluations, while the high-quality training set is larger at $N_\mathrm{HQ} = 2048$ to reduce the difference between the baselines of different runs. All experiments are repeated over five independent runs with different random seeds, and results are reported as the median.

A feature is dropped only once its weight falls below the current threshold, at which point the weight is set exactly to zero and the feature is removed from the representation. This threshold is initially inactive (set to zero). Early in the optimization the weights of irrelevant features decay quickly, however, as the optimization progresses, the optimizer settles into a minimum where the weights barely change and the active feature count stops decreasing. Because this minimum may only be local, we use it as a trigger: once such a plateau is detected, the threshold is raised to the 5$^\textrm{th}$ percentile of the remaining weights, dropping the least important of the surviving features, and each subsequent plateau cranks the threshold up further. If a deeper minimum exists, the optimizer then moves downhill toward it; if the plateau is already the best attainable, the procedure begins to remove genuinely relevant features, so the RMSE rises slowly until --- once the representation is over-compressed --- it returns to the unweighted baseline, now with far fewer features.

\section{Results}

\subsection{Property-dependent representations}
\label{subsec:pdr}

The weight vector is property-dependent: the optimization trajectory and the resulting weights encode which features are relevant for the target property, so different properties promote different features and suppress different ones. Figure~\ref{fig:cMBDF_QM9_weight_heatmap} illustrates this for cMBDF on QM9, where the two-body (radial) and three-body (angular) feature blocks are distinguished. The relative importance of the two blocks follows a physically interpretable progression across properties. The most glaring aspect is that out of 1880 features in total, 720 are irrelevant for all four considered properties, i.e. 38\,\% of the features can be dropped without loss of accuracy for any of them.

For the total energy, the largest contributions are pairwise interactions. Heat capacity reflects the curvature of the energy surface and involves both bond-stretching and angle-bending normal modes, leaving two-body and three-body features comparably important. The dipole moment depends on the vectorial sum of bond dipoles, which is governed by bond angles as much as by bond polarities, shifting the balance toward three-body features, but generally using fewer features of the full representation, since learning is inefficient as the representation poorly represents the dipole moment. Polarizability, as the response of the dipole to an external field, is even more sensitive to molecular shape and geometry, and shows the strongest three-body dominance of the four properties. The progression from total energy to polarizability thus tracks an increasing sensitivity to molecular geometry over pairwise distances.

The sorted weight profiles within each block (overlaid curves in Figure~\ref{fig:cMBDF_QM9_weight_heatmap}) reveal a consistent heavy-tail structure: a small number of features carry the bulk of the weight, followed by a long plateau of weakly contributing ones. This is consistent with the intrinsic dimensionality analysis of the same properties~\cite{Banjafar2025d}, which found that the property surface decomposes into a small set of separable degrees of freedom responsible for most of the variance, followed by a broad coupled regime. The optimization recovers an analogous structure in representation space, identifying the high-weight features as the effective separable directions for each property.

To quantify the property-specific component of the learned weights, we apply the total energy weight vector to the remaining three properties. This cross-property transfer still improves accuracy --- by 7.2\%, 1.3\%, and 7.6\% for heat capacity, dipole moment, and polarizability respectively --- indicating that a subset of features suppressed by the energy optimization is also irrelevant for other properties. However, these gains are substantially smaller than those achieved by property-specific optimization ($\sim$20\%, $\sim$10\%, and $\sim$15\%, observed in Figure~\ref{fig:qm9_compression}). The gap between cross-property and property-specific improvement directly quantifies the property-specific information encoded in the learned weights: the optimization identifies not only universally irrelevant features but also features that are selectively informative for each target property.

\subsection{Accuracy}
\label{subsec:accuracy}

\begin{figure*}
    \centering
    \includegraphics[width=\textwidth]{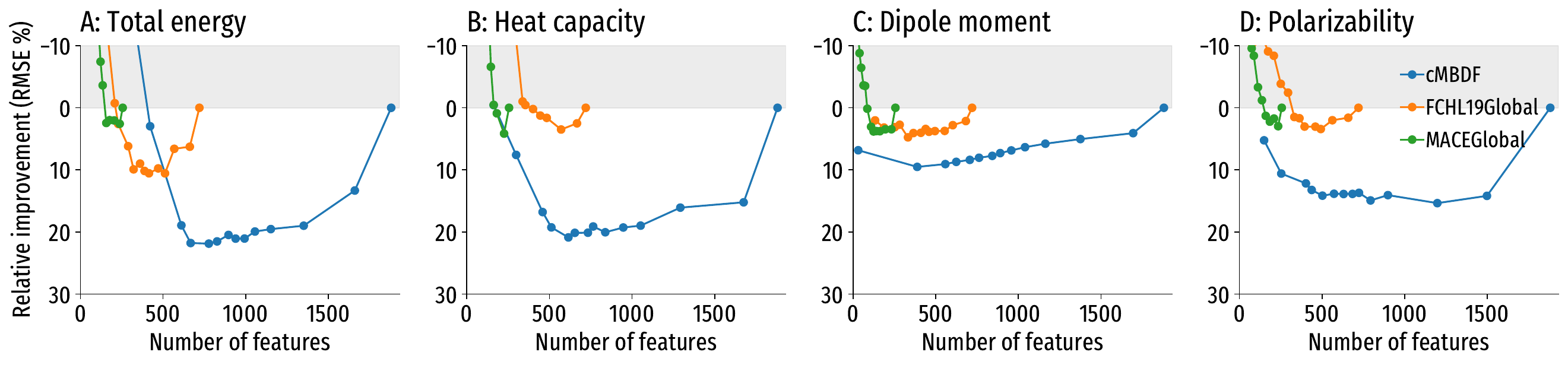}
    \caption{Compression curves showing relative RMSE improvement over the full-feature KRR baseline as a function of the number of retained features, for three global molecular representations on QM9. Each point is the median over five independent runs with different random seeds, tracing the optimization trajectory from full dimensionality (right) to maximum compression (left). The shaded region marks performance worse than the unweighted baseline.}
    \label{fig:qm9_compression}
\end{figure*}

Applying the optimized weight vector consistently improves KRR accuracy across all representations and properties on QM9. Figure~\ref{fig:qm9_compression} shows the relative RMSE improvement as a function of the number of retained features along the optimization trajectory; the rightmost point of each curve corresponds to the unweighted baseline, and any improvement is measured relative to it.

The largest gains are observed for cMBDF, with approximately 20\% RMSE reduction for total energy and heat capacity, $\sim$15\% for polarizability, and $\sim$10\% for dipole moment. FCHL19 shows consistent but more modest improvements, reaching $\sim$10\% for total energy and $\sim$5\% for the remaining properties. MACE improves least, with gains below 5\% in all cases.

This ranking reflects the inherent nature of each representation. cMBDF and FCHL19 are linear combinations of physics-inspired descriptors that enumerate physical interactions without reference to any specific target property, leaving a substantial fraction of features weakly relevant --- dropping these via weight optimization yields the largest accuracy gains. MACE, by contrast, couples physics-inspired features non-linearly as a foundation model pre-trained on a large and chemically diverse dataset.

Across all three representations, the dipole moment $\mu$ consistently yields the smallest improvement. This pattern is consistent with the known difficulty of learning $\mu$ from local atomic descriptors~\cite{Faber2017, Veit2020,Ko2021,Unke2019,CoolsCeuppens2022} (which may be partially improved by considering it as explicit response function\cite{Christensen_2019}): unlike energy or heat capacity, which are approximately additive over local atomic contributions, the dipole moment is a non-additive global response property governed by long-range charge distributions across the entire molecule. As a result, relevant features are expanded poorly in a two- and three-body basis where many features are required to observe learning at all, but none of them dominate. Feature weight optimization gains the most leverage when relevance is sparse and loses leverage when relevance is spread uniformly. This interpretation is consistent with the weight distributions in Fig.~\ref{fig:cMBDF_QM9_weight_heatmap}, where $\mu$ displays the least pronounced contrast between two-body and three-body features among the four properties.

\subsection{Compressibility}

A property-optimized representation is also a shorter one, which is helpful e.g. for encrypted model inference\cite{Weinreich2023a}. Through the optimization, features whose weights decay to zero are progressively removed, yielding a descriptor that grows more compact at each step. As shown in Fig.~\ref{fig:overview} (top right), irrelevant features decay toward zero and are eventually dropped, while the remaining features are promoted. Because the optimization targets the low quality data RMSE, it first eliminates irrelevant features  for both low and high quality data -- yielding accuracy gains in both. As optimization proceeds, however,  the process begins to overfit towards the low quality data, causing the high quality RMSE to increase. This bends the curves in Fig.~\ref{fig:qm9_compression} upward at low feature counts. The resulting bowl shape, observed across almost all representations and properties, defines the accuracy-compressibility tradeoff and identifies the optimal compression point for each case.

Figure~\ref{fig:qm9_compression} quantifies this tradeoff for all three representations. cMBDF, with 1880 features, shows the greatest absolute compressibility. For total energy, the best accuracy is achieved with 777 features, and the unweighted baseline accuracy is matched with fewer than 422 -- 22 percent of the full length of the descriptor. The other properties compress more aggressively: heat capacity reaches its optimum at 299 features, polarizability at 149, and dipole moment at only 17 features. The extreme case of dipole moment — where only 17 out of 1880 features are sufficient — reflects the same non-locality that limits accuracy improvement: cMBDF's local two- and three-body enumeration cannot encode the long-range charge distribution that governs $\mu$, so the vast majority of features contribute negligible predictive signal and can be discarded without penalty. The compression is large precisely because so little of the descriptor is useful, not because the retained subset is exceptionally informative.

A qualitatively similar pattern is observed for FCHL19 (720 features) and MACE (256 features). FCHL19 behaves analogously to cMBDF, as expected from their shared physics-based design: roughly two-thirds of features can be dropped across most properties. MACE presents a more striking result: despite yielding only modest accuracy improvements (see above), it shows larger compressibility across all properties Table~\ref{tab:compression}. Since MACE embeddings are extracted from a foundation model pre-trained on chemically diverse data, this redundancy reflects the mismatch between the broad generality of the pre-trained features and the narrow requirements of individual property regression. 

Beyond the magnitude of compression, the shape of the optimization curves reveals an important practical distinction between representations. cMBDF exhibits broad, flat bowl-shaped optima: for total energy, the improvement remains near its maximum across a range of roughly 300 to 700 features, meaning the compressed representation is robust to the stopping point of the optimization. FCHL19 and MACE, by contrast, show narrow, peaked optima — compression gains drop off sharply if too many or too few features are retained, requiring more precise control over the feature count. The dipole moment curve for cMBDF in Fig.~\ref{fig:qm9_compression}C illustrates this most clearly: the improvement varies only weakly from 17 features up to several hundred, confirming that the discarded features are genuinely redundant rather than mildly informative. This breadth reflects the greater intrinsic redundancy of cMBDF, consistent with the weight distributions discussed earlier.

\begin{figure*}[htbp]
    \centering
    \includegraphics[width=\textwidth]{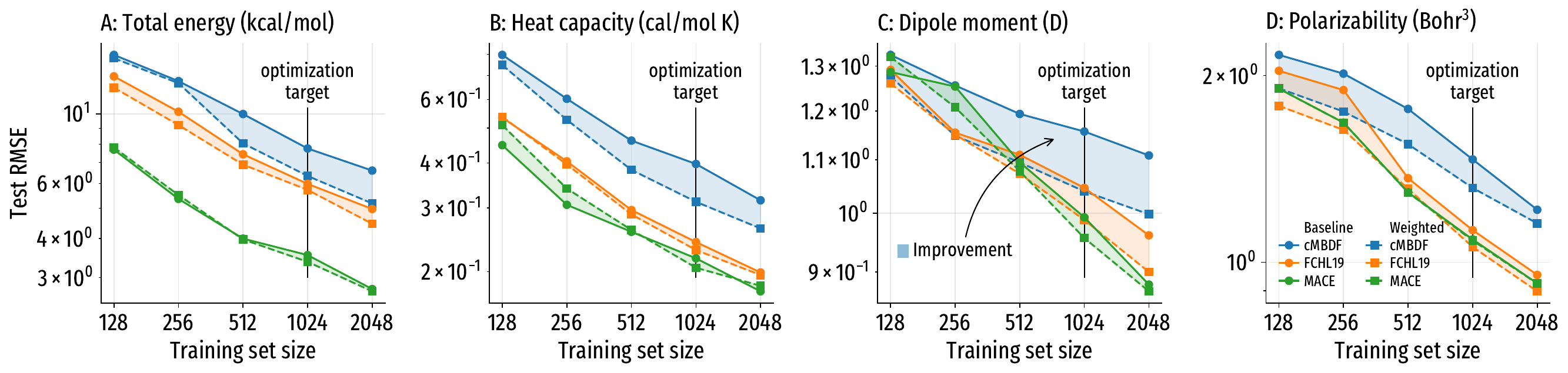}
    \caption{Learning curves comparing the weighted global representations (dashed) and the uncompressed baseline (solid) on QM9. Shaded areas: improvement through this work.}
    \label{fig:qm9_learning_curves}
\end{figure*}

Table~\ref{tab:compression} collects these observations into two complementary metrics: compression and data saved. The compression column measures how many dimensions of the representation can be dropped without loss of accuracy. The data saved column measures data efficiency: what fraction of high quality training points can be saved purely due to the compression. Concretely, it is the relative excess of training data the unweighted model would need -- read from its learning curve (Fig.~\ref{fig:qm9_learning_curves}) -- to match the accuracy the compressed model achieves at $N_\mathrm{HQ}=1024$. In many cases the two metrics reinforce each other: cMBDF on dipole moment achieves 98\% compression alongside 81\% data savings, and the same representation on total energy yields 84\% compression with 53\% savings. This coupling is intuitive -- a sparser, more targeted descriptor concentrates kernel similarity on structure genuinely predictive of the target property, possibly steepening the learning curve and reducing the number of high quality training points needed. However, compression and data savings can decouple: MACE on polarizability compresses by 47\% yet saves only 2\% of training data; cMBDF on polarizability achieves 92\% compression with only 32\% savings. In these cases the removed features are redundant for the kernel at the chosen evaluation point, but the retained features do not discriminate molecules sharply enough to meaningfully accelerate learning. The decoupling is most pronounced for MACE: if the compressed representation improves little over the baseline at fixed $N_\mathrm{HQ}$, the learning curve shifts correspondingly little, and data savings remain small even when nearly half the features are discarded.

\begin{table}[ht]
\centering
\caption{Feature compression and data efficiency improvement on QM9 for three representations
(cMBDF, FCHL19, MACE) across four molecular properties.
\textit{Compression}: percentage of features dropped at the point where the
weighted model test error first reaches the unweighted baseline.
\textit{Data saved}: percentage of training data rendered redundant by feature
weighting, measured as the fraction by which the equivalent unweighted
training set size exceeds $N=1024$.}
\label{tab:compression}
\begin{tabular}{l cccc cccc}
\toprule
& \multicolumn{4}{c}{Compression} & \multicolumn{4}{c}{Data saved} \\
\cmidrule(lr){2-5} \cmidrule(lr){6-9}
& $E$ & $C_v$ & $\mu$ & $\alpha$
& $E$ & $C_v$ & $\mu$ & $\alpha$ \\
\midrule
cMBDF  & 78\% & 91\% & 98\% & 92\% & 53\% & 49\% & 81\% & 32\% \\
FCHL19 & 71\% & 51\% & 82\% & 59\% & 15\% & 17\% & 37\% & 23\% \\
MACE   & 47\% & 36\% & 72\% & 47\% & 13\% & 18\% & 19\% &  2\% \\
\bottomrule
\end{tabular}
\end{table}

\subsection{Other dataset: VQM24}

\begin{figure}[htbp]
    \centering
    \includegraphics[width=\columnwidth]{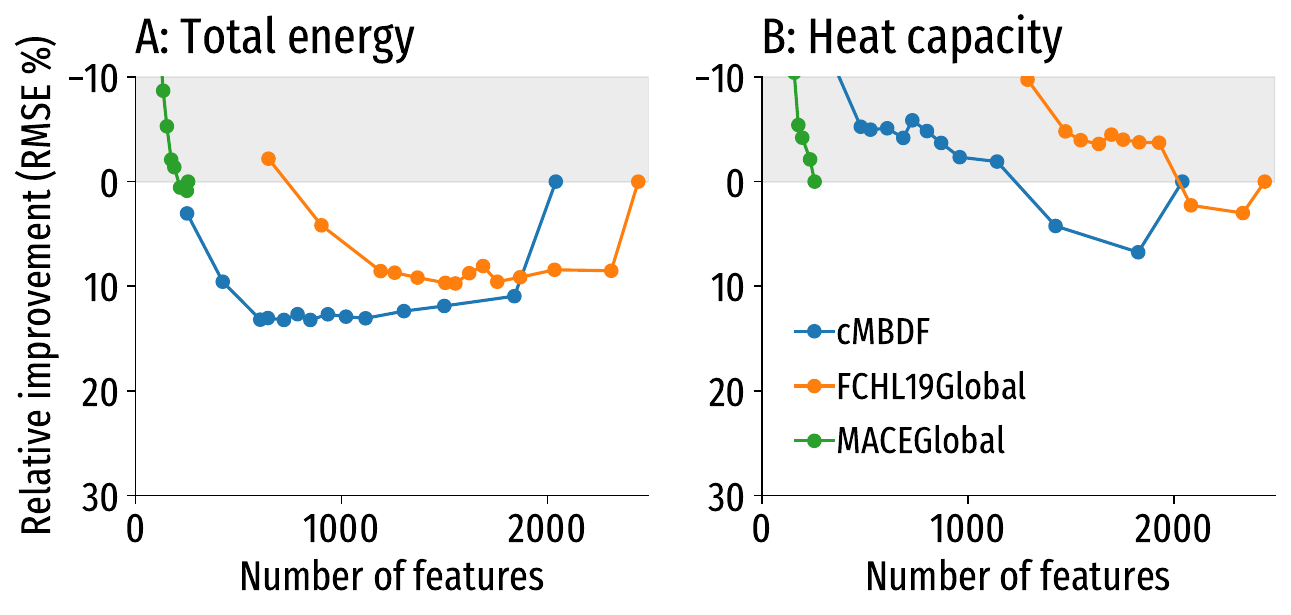}
    \caption{Compression curves for three global molecular representations on the larger and more chemically diverse VQM24 dataset, for two properties---total energy (A) and heat capacity (B). Axes, shading, and methodology are as in Fig.~\ref{fig:qm9_compression}.}
    \label{fig:vqm24_compression}
\end{figure}

Many representations have been built with QM9 as target benchmark set in mind, so it is relevant to consider other databases. We consider the total energy and heat capacity on the newer VQM24 dataset across all three representations (Fig.~\ref{fig:vqm24_compression}). VQM24 spans a broader chemical space, including a wider range of elements and molecular compositions, which makes KRR training less efficient and representations less able to encode molecular information compactly. The learning curves on VQM24 are noticeably flatter than on QM9, providing a more demanding test of whether the optimization generalizes.

For total energy, the method transfers successfully. cMBDF achieves around 13\% improvement at 849 features, with a very flat bowl spanning roughly 500 to 1800 features -- the compressed representation is robust to the stopping point of the optimization. FCHL19 shows a similar bowl-shaped behavior, reaching around 10\% improvement at 1553 features and matching baseline accuracy at 903 features. MACE shows no meaningful improvement and degrades rapidly upon compression, consistent with its behavior on QM9.

For heat capacity, the story is different. On QM9, heat capacity produced wide bowl-shaped curves with up to 20\% improvement for cMBDF; on VQM24, only the compression aspect remains, but only marginal accuracy gains remain: cMBDF peaks at only around 7\% near the full descriptor size and crosses baseline at approximately 1400 features, while FCHL19 reaches barely 3\% before immediately declining. Neither curve shows a bowl shape. This reflects the nature of VQM24: the greater chemical diversity distributes heat capacity information more uniformly across the descriptor, leaving little concentrated redundancy for the optimization to exploit. MACE shows no improvement and degrades immediately upon any compression.

Taken together, the VQM24 results confirm that the method generalizes across datasets, but its effectiveness scales with the available redundancy in the descriptor. Total energy, being approximately additive and well-encoded by local features, remains compressible even in a chemically diverse setting. Heat capacity, requiring a more distributed representation of molecular vibrations, proves resilient to compression when the training data spans a wide chemical space.

\begin{figure}[htbp]
    \centering
    \includegraphics[width=\columnwidth]{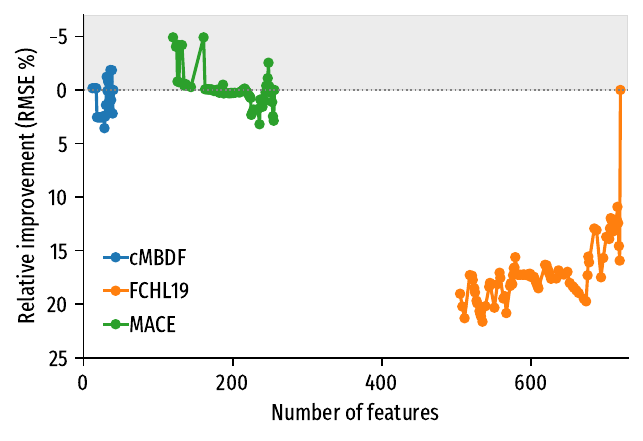}
    \caption{Compression curves for local representations on QM9 total energy. Per-atom descriptors are used directly with their respective optimal local kernels: the elemental kernel for cMBDF and FCHL19, and the all-pair local kernel for MACE. Axes and shading follow the same convention as Fig.~\ref{fig:qm9_compression}. }
    \label{fig:comparison_local}
\end{figure}

\subsection{Local representations}

The global compression results reported above aggregate per-atom contributions into a single molecular vector via summation. We now examine whether the same learning scheme extends to local representations, where per-atom descriptors are used directly and molecular similarity is computed via the atom-pair kernels.

Figure~\ref{fig:comparison_local} shows compression curves for QM9 total energy using local kernels: the elemental kernel for cMBDF and FCHL19, and the all-pair local kernel for MACE. The three representations behave very differently in this setting.

FCHL19 with the elemental kernel achieves the largest gain of any configuration tested: approximately 19--21\% RMSE reduction over a broad plateau spanning 500--700 features. This improvement matches or exceeds the global FCHL19 gain for total energy, and the wide optimum again indicates robustness to the stopping criterion.

cMBDF as local representation is already designed to be much shorter and on this dataset has 40 features, compared to $\sim$1880 in the global case. The optimization compresses this from 40 to  25 features with a $\sim$3--4\% improvement --- modest in absolute terms, but notable given that the starting representation is already compact. The smaller gain relative to global cMBDF is consistent with diminished redundancy at the per-atom level: the summation step in the global case introduces additional redundancy across atomic environments that feature weighting can exploit, but which is absent in the local descriptor.

MACE with the local kernel shows no improvement; the test error degrades immediately as features are removed. This is consistent with the global MACE behavior and reflects the compact nature of pre-trained embeddings: the 256-dimensional per-atom features encode structure relevant to total energy without the cross-atom redundancy that makes the global summed representation compressible.

\subsection{Learning curves -- practical payoff / data efficiency}

The compression curves above evaluate accuracy at a fixed high-quality training set size of $N_\mathrm{HQ} = 1024$. Fig.~\ref{fig:qm9_learning_curves} extends this picture by showing full learning curves for all three representations across all four QM9 properties, comparing the uncompressed baseline (solid lines) to the weighted model (dashed lines) at training sizes $N = 128$--$2048$, i.e. including data set sizes that have not been optimization targets. The weight vector is fixed to the one selected at $N_\mathrm{HQ} = 1024$; no re-optimization is performed at other training sizes. Two qualitatively distinct effects are visible: a \textit{vertical shift}, where weighted and baseline curves are roughly parallel (constant relative improvement across all $N$), and a \textit{slope change}, indicating that additional data can now be used more efficiently.

For total energy and heat capacity, the dominant effect is a vertical shift. The learning curve for the weighted version of FCHL19 and cMBDF run parallel to their respective baseline, indicating a stable accuracy and simultaneous efficiency gain at all training sizes. The cMBDF curves not only shifts downward but exhibits a modest steepening, with the compressed model with 512 points reaching accuracy comparable to the unweighted baseline at half of the data for both properties. The most striking effect appears for the dipole moment in Fig.~\ref{fig:qm9_learning_curves}C. The cMBDF baseline is comparably flat, i.e. cannot make efficient use of additional training data. Since the weighted version improves, this is likely because the full 1880-dimensional descriptor space contains too many entries which are unrelated for this property. The weighted cMBDF curve is substantially steeper: the compressed model at $N = 256$ already matches the accuracy of the unweighted baseline at more than four times the number of high-quality data points, which is the single largest absolute data-efficiency gain in the figure. The growing gap between the two curves confirms that the benefit of compression increases with number of data points, an effect that is also found for FCHL19, although less pronounced. For the polarizability in Fig.~\ref{fig:qm9_learning_curves}D, both cMBDF and FCHL19 benefit from the compression, typically by a constant factor which is most likely to come from the filtering of irrelevant features. 

In summary, the learning curves demonstrate that the benefit of transfer compression is not confined to the evaluation point used during optimization. For properties with concentrated feature relevance (energy, heat capacity), compression yields a uniform data efficiency gain. For properties where the full descriptor is nearly uninformative (dipole moment), compression can change the fundamental scaling of the learning curve, converting a data-limited, high-dimensional regression problem into a tractable low-dimensional one.

\section{Conclusion}

Our results show that chemical representations are over-complete, and that representations are achievable which at the same time are more accurate and shorter. This is done through transfer learning in feature space rather than in labels over two datasets of different fidelity where the low-quality dataset is used to extract the physically relevant subset of features only. We confirm the assumption that feature relevance is transferable between data computed at different levels of theory: since a weight vector optimized on low-quality data also works on high-quality data, feature relevance proves transferable for the four physical properties studied and for different datasets, representations and both global and local kernels.

In particular, for representations built on systematically enumerated physical many-body terms, such as cMBDF and FCHL19, the number of irrelevant features is more pronounced. In addition, we observe weighting features higher that originally contribute weakly to the feature space has a strong effect. By analyzing the cMBDF weight vector, we further observe that the final weights agree with the physical interpretation for each target property -- e.g., two-body features play a more important role for total energy, whereas for polarizability the three-body terms dominate.

We show that shorter and more accurate representations are not limited to a single dataset or representation flavor (local and global). By optimizing the weight vector on VQM24 -- a larger and more chemically diverse set of molecules -- we find that representations which perform fairly poorly on this harder dataset can still be improved through a property-dependent representation. The method also extends to local representations, which are typically more accurate and more compact than their global counterparts for extensive properties. The accuracy gains here are representation-dependent -- FCHL19 reaches its largest improvement of all in the local setting, while for cMBDF the local gain is more modest than the global -- but the compressibility keeps its impact: even the very short local cMBDF (40 features) can be compressed further. The pretrained representation MACE leads in accuracy but saw large numbers of  high quality training data (albeit for materials not molecules) during representation design. Our method can even compress the non-linear neural network descriptor of MACE, but achieves only modest accuracy improvement. We expect that for stronger improvements, compression would need to be done at training time for neural network representations.

Overall, this work provides a method to construct property-dependent chemical representations from more general ones, points out inefficiencies, and demonstrates that representations can be shorter and more accurate at the same time. Moreover, our method doubles as diagnostic tool to detect whether a given set of features is expressive enough for a given property, since the initial weighting during representation design is somewhat arbitrary and typically chosen to work well on the benchmark properties considered.

Main limitations for our method are that it assumes that a sufficiently predictive but cheap baseline such as xTB exists for the property in question and that the initial representation needs to be short enough such that optimization is feasible. Furthermore, global representations are much preferred as their optimization is orders of magnitude cheaper and less noisy than the one for local representations. 

On a practical side, shorter representations drastically reduce cost e.g. for encrypted inference\cite{Weinreich2023a} or reduce the high memory requirements for local representations and the associated compute time which dominates KRR training cost\cite{Grimblat2026}. Employing the filtering during representation design may be beneficial to select effective feature spaces for elusive properties -- both for kernel methods or neural networks. Possible extensions could be to find representations for different regions in chemical space, or to steer active learning selections towards data points which span the selected feature space more dominantly.

\bibliography{nablachem,add} 

@article{Grimblat2026,
  title = {Data-Driven Complete Basis Set Limit Estimates from a Minimal Auxiliary Basis},
  author = {Grimblat, Nicolas and Klassen, Gabriel and {von Rudorff}, Guido Falk},
  year = 2026,
  publisher = {arXiv},
  doi = {10.48550/ARXIV.2605.15927},
  journal = {arXiv},
  urldate = {2026-05-27},
  copyright = {arXiv.org perpetual, non-exclusive license}
}

@misc{Khan2024,
	title = {Towards comprehensive coverage of chemical space: {Quantum} mechanical properties of 836k constitutional and conformational closed shell neutral isomers consisting of {HCNOFSiPSClBr}},
	doi = {10.48550/ARXIV.2405.05961},
	publisher = {arXiv},
	author = {Khan, Danish and Benali, Anouar and Kim, Scott Y. H. and von Rudorff, Guido Falk and von Lilienfeld, O. Anatole},
	year = {2024},
}

@article{Ruddigkeit2012,
	title = {Enumeration of 166 billion organic small molecules in the chemical universe database {GDB}-17},
	volume = {52},
	doi = {10.1021/ci300415d},
	number = {11},
	journal = {Journal of Chemical Information and Modeling},
	publisher = {American Chemical Society (ACS)},
	author = {Ruddigkeit, Lars and van Deursen, Ruud and Blum, Lorenz C. and Reymond, Jean-Louis},
	month = nov,
	year = {2012},
	pages = {2864--2875},
}

@article{Ramakrishnan2014,
	title = {Quantum chemistry structures and properties of 134 kilo molecules},
	volume = {1},
	doi = {10.1038/sdata.2014.22},
	journal = {Scientific Data},
	publisher = {Springer Nature},
	author = {Ramakrishnan, Raghunathan and Dral, Pavlo O. and Rupp, Matthias and von Lilienfeld, O. Anatole},
	month = aug,
	year = {2014},
}

@article{vonLillienfeld2018,
	title = {Quantum machine learning in chemical compound space},
	volume = {57},
	doi = {10.1002/anie.201709686},
	number = {16},
	journal = {Angewandte Chemie International Edition},
	publisher = {Wiley},
	author = {von Lilienfeld, O. Anatole},
	month = mar,
	year = {2018},
	pages = {4164--4169},
}

@article{Montavon2013,
	title = {Machine learning of molecular electronic properties in chemical compound space},
	volume = {15},
	doi = {10.1088/1367-2630/15/9/095003},
	number = {9},
	journal = {New Journal of Physics},
	publisher = {IOP Publishing},
	author = {Montavon, Grégoire and Rupp, Matthias and Gobre, Vivekanand and Vazquez-Mayagoitia, Alvaro and Hansen, Katja and Tkatchenko, Alexandre and Müller, Klaus-Robert and von Lilienfeld, O Anatole},
	month = sep,
	year = {2013},
	pages = {095003},
}

@article{Ramakrishnan2015,
	title = {Big data meets quantum chemistry approximations: {The} Δ-machine learning approach},
	volume = {11},
	url = {https://doi.org/10.1021/acs.jctc.5b00099},
	doi = {10.1021/acs.jctc.5b00099},
	number = {5},
	journal = {Journal of Chemical Theory and Computation},
	author = {Ramakrishnan, Raghunathan and Dral, Pavlo O. and Rupp, Matthias and von Lilienfeld, O. Anatole},
	year = {2015},
	pages = {2087--2096},
}

@article{Christensen_2019,
	title = {Operator quantum machine learning: {Navigating} the chemical space of response properties},
	volume = {73},
	doi = {10.2533/chimia.2019.1028},
	number = {12},
	journal = {CHIMIA International Journal for Chemistry},
	publisher = {Swiss Chemical Society},
	author = {Christensen, Anders S. and von Lilienfeld, O. Anatole},
	month = dec,
	year = {2019},
	pages = {1028--1031},
}

@article{Faber2018,
	title = {Alchemical and structural distribution based representation for universal quantum machine learning},
	volume = {148},
	doi = {10.1063/1.5020710},
	number = {24},
	journal = {The Journal of Chemical Physics},
	publisher = {AIP Publishing},
	author = {Faber, Felix A. and Christensen, Anders S. and Huang, Bing and von Lilienfeld, O. Anatole},
	month = jun,
	year = {2018},
	pages = {241717},
}

@article{Zaspel2018,
	title = {Boosting quantum machine learning models with a multilevel combination technique: {Pople} diagrams revisited},
	volume = {15},
	doi = {10.1021/acs.jctc.8b00832},
	number = {3},
	journal = {Journal of Chemical Theory and Computation},
	publisher = {American Chemical Society (ACS)},
	author = {Zaspel, Peter and Huang, Bing and Harbrecht, Helmut and von Lilienfeld, O. Anatole},
	month = dec,
	year = {2018},
	pages = {1546--1559},
}

@article{Behler2021,
	title = {Four generations of high-dimensional neural network potentials},
	volume = {121},
	doi = {10.1021/acs.chemrev.0c00868},
	number = {16},
	journal = {Chemical Reviews},
	publisher = {American Chemical Society (ACS)},
	author = {Behler, Jörg},
	month = mar,
	year = {2021},
	pages = {10037--10072},
}

@article{Husch2021,
	title = {Improved accuracy and transferability of molecular-orbital-based machine learning: {Organics}, transition-metal complexes, non-covalent interactions, and transition states},
	volume = {154},
	doi = {10.1063/5.0032362},
	number = {6},
	journal = {The Journal of Chemical Physics},
	publisher = {AIP Publishing},
	author = {Husch, Tamara and Sun, Jiace and Cheng, Lixue and Lee, Sebastian J. R. and Miller, Thomas F.},
	month = feb,
	year = {2021},
	pages = {064108},
}

@article{CoolsCeuppens2022,
	title = {Modeling electronic response properties with an explicit-electron machine learning potential},
	volume = {18},
	doi = {10.1021/acs.jctc.1c00978},
	number = {3},
	journal = {Journal of Chemical Theory and Computation},
	publisher = {American Chemical Society (ACS)},
	author = {Cools-Ceuppens, Maarten and Dambre, Joni and Verstraelen, Toon},
	month = feb,
	year = {2022},
	pages = {1672--1691},
}

@article{Welborn2018,
	title = {Transferability in machine learning for electronic structure via the molecular orbital basis},
	volume = {14},
	doi = {10.1021/acs.jctc.8b00636},
	number = {9},
	journal = {Journal of Chemical Theory and Computation},
	publisher = {American Chemical Society (ACS)},
	author = {Welborn, Matthew and Cheng, Lixue and Miller, Thomas F.},
	month = jul,
	year = {2018},
	pages = {4772--4779},
}

@article{Ko2021,
	title = {A fourth-generation high-dimensional neural network potential with accurate electrostatics including non-local charge transfer},
	volume = {12},
	doi = {10.1038/s41467-020-20427-2},
	number = {1},
	journal = {Nature Communications},
	publisher = {Springer Science and Business Media LLC},
	author = {Ko, Tsz Wai and Finkler, Jonas A. and Goedecker, Stefan and Behler, Jörg},
	month = jan,
	year = {2021},
}

@article{Smith2019,
	title = {Approaching coupled cluster accuracy with a general-purpose neural network potential through transfer learning},
	volume = {10},
	doi = {10.1038/s41467-019-10827-4},
	number = {1},
	journal = {Nature Communications},
	publisher = {Springer Science and Business Media LLC},
	author = {Smith, Justin S. and Nebgen, Benjamin T. and Zubatyuk, Roman and Lubbers, Nicholas and Devereux, Christian and Barros, Kipton and Tretiak, Sergei and Isayev, Olexandr and Roitberg, Adrian E.},
	month = jul,
	year = {2019},
}

@article{Weinreich2023a,
	title = {Encrypted machine learning of molecular quantum properties},
	volume = {4},
	doi = {10.1088/2632-2153/acc928},
	number = {2},
	journal = {Machine Learning: Science and Technology},
	publisher = {IOP Publishing},
	author = {Weinreich, Jan and von Rudorff, Guido Falk and von Lilienfeld, O Anatole},
	month = apr,
	year = {2023},
	pages = {025017},
}

@article{Drautz2019,
	title = {Atomic cluster expansion for accurate and transferable interatomic potentials},
	volume = {99},
	issn = {2469-9969},
	doi = {10.1103/physrevb.99.014104},
	number = {1},
	journal = {Physical Review B},
	publisher = {American Physical Society (APS)},
	author = {Drautz, Ralf},
	month = jan,
	year = {2019},
	pages = {014104},
}

@article{Nelson2013,
	title = {Compressive sensing as a paradigm for building physics models},
	volume = {87},
	issn = {1550-235X},
	doi = {10.1103/physrevb.87.035125},
	number = {3},
	journal = {Physical Review B},
	publisher = {American Physical Society (APS)},
	author = {Nelson, Lance J. and Hart, Gus L. W. and Zhou, Fei and Ozoliņš, Vidvuds},
	month = jan,
	year = {2013},
	pages = {035125},
}

@article{Faber2017,
	title = {Prediction {Errors} of {Molecular} {Machine} {Learning} {Models} {Lower} than {Hybrid} {DFT} {Error}},
	volume = {13},
	issn = {1549-9618},
	url = {https://doi.org/10.1021/acs.jctc.7b00577},
	doi = {10.1021/acs.jctc.7b00577},
	number = {11},
	urldate = {2025-05-09},
	journal = {Journal of Chemical Theory and Computation},
	publisher = {American Chemical Society},
	author = {Faber, Felix A. and Hutchison, Luke and Huang, Bing and Gilmer, Justin and Schoenholz, Samuel S. and Dahl, George E. and Vinyals, Oriol and Kearnes, Steven and Riley, Patrick F. and von Lilienfeld, O. Anatole},
	month = nov,
	year = {2017},
	pages = {5255--5264},
}

@article{Thant2025,
	title = {Kernel regression methods for prediction of materials properties: {Recent} developments},
	volume = {6},
	issn = {2688-4070},
	shorttitle = {Kernel regression methods for prediction of materials properties},
	url = {https://pubs.aip.org/cpr/article/6/1/011306/3335893/Kernel-regression-methods-for-prediction-of},
	doi = {10.1063/5.0242118},
	number = {1},
	urldate = {2025-05-18},
	journal = {Chemical Physics Reviews},
	author = {Thant, Ye Min and Wakamiya, Taishiro and Nukunudompanich, Methawee and Kameda, Keisuke and Ihara, Manabu and Manzhos, Sergei},
	month = mar,
	year = {2025},
	pages = {011306},
}

@article{Shawe-Taylor2005,
	title = {On the {Eigenspectrum} of the {Gram} {Matrix} and the {Generalization} {Error} of {Kernel}-{PCA}},
	volume = {51},
	copyright = {https://ieeexplore.ieee.org/Xplorehelp/downloads/license-information/IEEE.html},
	issn = {0018-9448},
	url = {http://ieeexplore.ieee.org/document/1459055/},
	doi = {10.1109/TIT.2005.850052},
	number = {7},
	urldate = {2025-05-21},
	journal = {IEEE Transactions on Information Theory},
	author = {Shawe-Taylor, J. and Williams, C.K.I. and Cristianini, N. and Kandola, J.},
	month = jul,
	year = {2005},
	pages = {2510--2522},
}

@article{Khan2023a,
	title = {Kernel based quantum machine learning at record rate: {Many}-body distribution functionals as compact representations},
	volume = {159},
	issn = {0021-9606, 1089-7690},
	shorttitle = {Kernel based quantum machine learning at record rate},
	url = {https://pubs.aip.org/jcp/article/159/3/034106/2902959/Kernel-based-quantum-machine-learning-at-record},
	doi = {10.1063/5.0152215},
	number = {3},
	urldate = {2025-06-10},
	journal = {The Journal of Chemical Physics},
	author = {Khan, Danish and Heinen, Stefan and Von Lilienfeld, O. Anatole},
	month = jul,
	year = {2023},
	pages = {034106},
}

@article{Nandi2023,
	title = {{MultiXC}-{QM9}: {Large} dataset of molecular and reaction energies from multi-level quantum chemical methods},
	volume = {10},
	issn = {2052-4463},
	shorttitle = {{MultiXC}-{QM9}},
	url = {https://www.nature.com/articles/s41597-023-02690-2},
	doi = {10.1038/s41597-023-02690-2},
	number = {1},
	urldate = {2025-06-03},
	journal = {Scientific Data},
	author = {Nandi, Surajit and Vegge, Tejs and Bhowmik, Arghya},
	month = nov,
	year = {2023},
	pages = {783},
}

@article{Karandashev2022,
	title = {An orbital-based representation for accurate quantum machine learning},
	volume = {156},
	issn = {0021-9606, 1089-7690},
	url = {https://pubs.aip.org/jcp/article/156/11/114101/2840879/An-orbital-based-representation-for-accurate},
	doi = {10.1063/5.0083301},
	number = {11},
	urldate = {2025-06-10},
	journal = {The Journal of Chemical Physics},
	author = {Karandashev, Konstantin and Von Lilienfeld, O. Anatole},
	month = mar,
	year = {2022},
	pages = {114101},
}

@article{Musil2021,
	title = {Physics-{Inspired} {Structural} {Representations} for {Molecules} and {Materials}},
	volume = {121},
	copyright = {https://creativecommons.org/licenses/by/4.0/},
	issn = {0009-2665, 1520-6890},
	url = {https://pubs.acs.org/doi/10.1021/acs.chemrev.1c00021},
	doi = {10.1021/acs.chemrev.1c00021},
	number = {16},
	urldate = {2025-06-10},
	journal = {Chemical Reviews},
	author = {Musil, Felix and Grisafi, Andrea and Bartók, Albert P. and Ortner, Christoph and Csányi, Gábor and Ceriotti, Michele},
	month = aug,
	year = {2021},
	pages = {9759--9815},
}

@article{VonLilienfeld2020,
	title = {Exploring chemical compound space with quantum-based machine learning},
	volume = {4},
	issn = {2397-3358},
	url = {https://www.nature.com/articles/s41570-020-0189-9},
	doi = {10.1038/s41570-020-0189-9},
	number = {7},
	urldate = {2025-06-10},
	journal = {Nature Reviews Chemistry},
	author = {Von Lilienfeld, O. Anatole and Müller, Klaus-Robert and Tkatchenko, Alexandre},
	month = jun,
	year = {2020},
	pages = {347--358},
}

@article{Facco2017,
	title = {Estimating the intrinsic dimension of datasets by a minimal neighborhood information},
	volume = {7},
	issn = {2045-2322},
	url = {https://www.nature.com/articles/s41598-017-11873-y},
	doi = {10.1038/s41598-017-11873-y},
	number = {1},
	urldate = {2025-06-17},
	journal = {Scientific Reports},
	author = {Facco, Elena and d’Errico, Maria and Rodriguez, Alex and Laio, Alessandro},
	month = sep,
	year = {2017},
	pages = {12140},
}

@article{Fedik2022,
	title = {Extending machine learning beyond interatomic potentials for predicting molecular properties},
	volume = {6},
	issn = {2397-3358},
	url = {https://www.nature.com/articles/s41570-022-00416-3},
	doi = {10.1038/s41570-022-00416-3},
	number = {9},
	urldate = {2025-06-17},
	journal = {Nature Reviews Chemistry},
	author = {Fedik, Nikita and Zubatyuk, Roman and Kulichenko, Maksim and Lubbers, Nicholas and Smith, Justin S. and Nebgen, Benjamin and Messerly, Richard and Li, Ying Wai and Boldyrev, Alexander I. and Barros, Kipton and Isayev, Olexandr and Tretiak, Sergei},
	month = aug,
	year = {2022},
	pages = {653--672},
}

@article{Stuke2019,
	title = {Chemical diversity in molecular orbital energy predictions with kernel ridge regression},
	volume = {150},
	issn = {0021-9606, 1089-7690},
	url = {https://pubs.aip.org/jcp/article/150/20/204121/199068/Chemical-diversity-in-molecular-orbital-energy},
	doi = {10.1063/1.5086105},
	number = {20},
	urldate = {2025-07-28},
	journal = {The Journal of Chemical Physics},
	publisher = {AIP Publishing},
	author = {Stuke, Annika and Todorović, Milica and Rupp, Matthias and Kunkel, Christian and Ghosh, Kunal and Himanen, Lauri and Rinke, Patrick},
	month = may,
	year = {2019},
}

@article{Fabregat2022,
	title = {Metric learning for kernel ridge regression: assessment of molecular similarity},
	volume = {3},
	issn = {2632-2153},
	shorttitle = {Metric learning for kernel ridge regression},
	url = {https://iopscience.iop.org/article/10.1088/2632-2153/ac8e4f},
	doi = {10.1088/2632-2153/ac8e4f},
	number = {3},
	urldate = {2025-11-01},
	journal = {Machine Learning: Science and Technology},
	author = {Fabregat, Raimon and Van Gerwen, Puck and Haeberle, Matthieu and Eisenbrand, Friedrich and Corminboeuf, Clémence},
	month = sep,
	year = {2022},
	pages = {035015},
}

@article{Willatt2018,
	title = {Feature optimization for atomistic machine learning yields a data-driven construction of the periodic table of the elements},
	volume = {20},
	issn = {1463-9076, 1463-9084},
	url = {https://xlink.rsc.org/?DOI=C8CP05921G},
	doi = {10.1039/C8CP05921G},
	number = {47},
	urldate = {2025-11-01},
	journal = {Physical Chemistry Chemical Physics},
	author = {Willatt, Michael J. and Musil, Félix and Ceriotti, Michele},
	year = {2018},
	pages = {29661--29668},
}

@article{Banjafar2025d,
	title = {Intrinsic dimensionality of molecular properties},
	volume = {163},
	issn = {0021-9606, 1089-7690},
	url = {https://pubs.aip.org/jcp/article/163/17/174301/3370563/Intrinsic-dimensionality-of-molecular-properties},
	doi = {10.1063/5.0289088},
	number = {17},
	urldate = {2025-11-06},
	journal = {The Journal of Chemical Physics},
	author = {Banjafar, Ali and Von Rudorff, Guido Falk},
	month = nov,
	year = {2025},
	pages = {174301},
}

@article{Huang2016,
	title = {Communication: {Understanding} molecular representations in machine learning: {The} role of uniqueness and target similarity},
	volume = {145},
	issn = {0021-9606},
	shorttitle = {Communication},
	url = {https://doi.org/10.1063/1.4964627},
	doi = {10.1063/1.4964627},
	number = {16},
	urldate = {2026-02-24},
	journal = {The Journal of Chemical Physics},
	author = {Huang, Bing and von Lilienfeld, O. Anatole},
	month = oct,
	year = {2016},
	pages = {161102},
}

@article{Khan2025a,
	title = {Generalized convolutional many-body distribution functional representations},
	volume = {122},
	url = {https://www.pnas.org/doi/abs/10.1073/pnas.2415662122},
	doi = {10.1073/pnas.2415662122},
	number = {41},
	urldate = {2026-02-24},
	journal = {Proceedings of the National Academy of Sciences},
	publisher = {Proceedings of the National Academy of Sciences},
	author = {Khan, Danish and von Lilienfeld, O. Anatole},
	month = oct,
	year = {2025},
	pages = {e2415662122},
}

@article{Schutt2018,
	title = {{SchNet} - a deep learning architecture for molecules and materials},
	volume = {148},
	issn = {0021-9606, 1089-7690},
	url = {http://arxiv.org/abs/1712.06113},
	doi = {10.1063/1.5019779},
	number = {24},
	urldate = {2026-02-24},
	journal = {The Journal of Chemical Physics},
	author = {Schütt, Kristof T. and Sauceda, Huziel E. and Kindermans, Pieter-Jan and Tkatchenko, Alexandre and Müller, Klaus-Robert},
	month = jun,
	year = {2018},
	pages = {241722},
}

@article{Unke2019,
	title = {{PhysNet}: {A} {Neural} {Network} for {Predicting} {Energies}, {Forces}, {Dipole} {Moments}, and {Partial} {Charges}},
	volume = {15},
	issn = {1549-9618},
	shorttitle = {{PhysNet}},
	url = {https://doi.org/10.1021/acs.jctc.9b00181},
	doi = {10.1021/acs.jctc.9b00181},
	number = {6},
	urldate = {2026-02-20},
	journal = {Journal of Chemical Theory and Computation},
	publisher = {American Chemical Society},
	author = {Unke, Oliver T. and Meuwly, Markus},
	month = jun,
	year = {2019},
	pages = {3678--3693},
}

@article{Batzner2022,
	title = {E(3)-equivariant graph neural networks for data-efficient and accurate interatomic potentials},
	volume = {13},
	copyright = {2022 The Author(s)},
	issn = {2041-1723},
	url = {https://www.nature.com/articles/s41467-022-29939-5},
	doi = {10.1038/s41467-022-29939-5},
	number = {1},
	urldate = {2026-02-20},
	journal = {Nature Communications},
	publisher = {Nature Publishing Group},
	author = {Batzner, Simon and Musaelian, Albert and Sun, Lixin and Geiger, Mario and Mailoa, Jonathan P. and Kornbluth, Mordechai and Molinari, Nicola and Smidt, Tess E. and Kozinsky, Boris},
	month = may,
	year = {2022},
	pages = {2453},
}

@article{Uhrin2021,
	title = {Through the eyes of a descriptor: {Constructing} complete, invertible descriptions of atomic environments},
	volume = {104},
	shorttitle = {Through the eyes of a descriptor},
	url = {https://link.aps.org/doi/10.1103/PhysRevB.104.144110},
	doi = {10.1103/PhysRevB.104.144110},
	number = {14},
	urldate = {2026-02-19},
	journal = {Physical Review B},
	publisher = {American Physical Society},
	author = {Uhrin, Martin},
	month = oct,
	year = {2021},
	pages = {144110},
}

@article{Vinod2023,
	title = {Multifidelity {Machine} {Learning} for {Molecular} {Excitation} {Energies}},
	volume = {19},
	copyright = {https://doi.org/10.15223/policy-029},
	issn = {1549-9618, 1549-9626},
	url = {https://pubs.acs.org/doi/10.1021/acs.jctc.3c00882},
	doi = {10.1021/acs.jctc.3c00882},
	number = {21},
	urldate = {2026-05-08},
	journal = {Journal of Chemical Theory and Computation},
	author = {Vinod, Vivin and Maity, Sayan and Zaspel, Peter and Kleinekathöfer, Ulrich},
	month = nov,
	year = {2023},
	pages = {7658--7670},
}

@article{Vinod2025,
	title = {Predicting {Molecular} {Energies} of {Small} {Organic} {Molecules} {With} {Multi}‐{Fidelity} {Methods}},
	volume = {46},
	issn = {0192-8651, 1096-987X},
	url = {https://onlinelibrary.wiley.com/doi/10.1002/jcc.70056},
	doi = {10.1002/jcc.70056},
	number = {6},
	urldate = {2026-05-08},
	journal = {Journal of Computational Chemistry},
	author = {Vinod, Vivin and Lyu, Dongyu and Ruth, Marcel and R. Schreiner, Peter and Kleinekathöfer, Ulrich and Zaspel, Peter},
	month = mar,
	year = {2025},
	pages = {e70056},
}

@incollection{MacKay1996,
	address = {Dordrecht},
	title = {Bayesian {Non}-{Linear} {Modeling} for the {Prediction} {Competition}},
	isbn = {978-90-481-4407-5 978-94-015-8729-7},
	url = {http://link.springer.com/10.1007/978-94-015-8729-7_18},
	doi = {10.1007/978-94-015-8729-7_18},
	urldate = {2026-06-01},
	booktitle = {Maximum {Entropy} and {Bayesian} {Methods}},
	publisher = {Springer Netherlands},
	author = {MacKay, David J. C.},
	editor = {Heidbreder, Glenn R.},
	year = {1996},
	pages = {221--234},
}

@article{Pozdnyakov2020,
	title = {Incompleteness of {Atomic} {Structure} {Representations}},
	volume = {125},
	issn = {0031-9007, 1079-7114},
	url = {https://link.aps.org/doi/10.1103/PhysRevLett.125.166001},
	doi = {10.1103/PhysRevLett.125.166001},
	number = {16},
	urldate = {2026-06-01},
	journal = {Physical Review Letters},
	author = {Pozdnyakov, Sergey N. and Willatt, Michael J. and Bartók, Albert P. and Ortner, Christoph and Csányi, Gábor and Ceriotti, Michele},
	month = oct,
	year = {2020},
	pages = {166001},
}

@article{Behler2007,
	title = {Generalized {Neural}-{Network} {Representation} of {High}-{Dimensional} {Potential}-{Energy} {Surfaces}},
	volume = {98},
	copyright = {http://link.aps.org/licenses/aps-default-license},
	issn = {0031-9007, 1079-7114},
	url = {https://link.aps.org/doi/10.1103/PhysRevLett.98.146401},
	doi = {10.1103/PhysRevLett.98.146401},
	number = {14},
	urldate = {2026-06-01},
	journal = {Physical Review Letters},
	author = {Behler, Jörg and Parrinello, Michele},
	month = apr,
	year = {2007},
	pages = {146401},
}

@article{Bartok2013,
	title = {On representing chemical environments},
	volume = {87},
	copyright = {http://link.aps.org/licenses/aps-default-license},
	issn = {1098-0121, 1550-235X},
	url = {https://link.aps.org/doi/10.1103/PhysRevB.87.184115},
	doi = {10.1103/PhysRevB.87.184115},
	number = {18},
	urldate = {2026-06-01},
	journal = {Physical Review B},
	author = {Bartók, Albert P. and Kondor, Risi and Csányi, Gábor},
	month = may,
	year = {2013},
	pages = {184115},
}

@article{Christensen2020a,
	title = {{FCHL} revisited: {Faster} and more accurate quantum machine learning},
	volume = {152},
	issn = {0021-9606, 1089-7690},
	shorttitle = {{FCHL} revisited},
	url = {https://pubs.aip.org/jcp/article/152/4/044107/1064737/FCHL-revisited-Faster-and-more-accurate-quantum},
	doi = {10.1063/1.5126701},
	number = {4},
	urldate = {2026-06-01},
	journal = {The Journal of Chemical Physics},
	author = {Christensen, Anders S. and Bratholm, Lars A. and Faber, Felix A. and Anatole Von Lilienfeld, O.},
	month = jan,
	year = {2020},
	pages = {044107},
}

@article{Batatia,
	title = {{MACE}: {Higher} {Order} {Equivariant} {Message} {Passing} {Neural} {Networks} for {Fast} and {Accurate} {Force} {Fields}},
	journal = {NeurIPS 35, arXiv 2206.07697},
	author = {Batatia, Ilyes and Kovács, Dávid Péter and Simm, Gregor N C and Ortner, Christoph and Csányi, Gábor},
	year = {2022},
}

@misc{Batatia2024,
	title = {A foundation model for atomistic materials chemistry},
	copyright = {Creative Commons Attribution Non Commercial No Derivatives 4.0 International},
	url = {https://arxiv.org/abs/2401.00096},
	doi = {10.48550/ARXIV.2401.00096},
	urldate = {2026-06-01},
	publisher = {arXiv},
	author = {Batatia, Ilyes and Benner, Philipp and Chiang, Yuan and Elena, Alin M. and Kovács, Dávid P. and Riebesell, Janosh and Advincula, Xavier R. and Asta, Mark and Avaylon, Matthew and Baldwin, William J. and Berger, Fabian and Bernstein, Noam and Bhowmik, Arghya and Bigi, Filippo and Blau, Samuel M. and Cărare, Vlad and Ceriotti, Michele and Chong, Sanggyu and Darby, James P. and De, Sandip and Della Pia, Flaviano and Deringer, Volker L. and Elijošius, Rokas and El-Machachi, Zakariya and Falcioni, Fabio and Fako, Edvin and Ferrari, Andrea C. and Gardner, John L. A. and Gawkowski, Mikolaj J. and Genreith-Schriever, Annalena and George, Janine and Goodall, Rhys E. A. and Grandel, Jonas and Grey, Clare P. and Grigorev, Petr and Han, Shuang and Handley, Will and Heenen, Hendrik H. and Hermansson, Kersti and Holm, Christian and Ho, Cheuk Hin and Hofmann, Stephan and Jaafar, Jad and Jakob, Konstantin S. and Jung, Hyunwook and Kapil, Venkat and Kaplan, Aaron D. and Karimitari, Nima and Kermode, James R. and Kourtis, Panagiotis and Kroupa, Namu and Kullgren, Jolla and Kuner, Matthew C. and Kuryla, Domantas and Liepuoniute, Guoda and Lin, Chen and Margraf, Johannes T. and Magdău, Ioan-Bogdan and Michaelides, Angelos and Moore, J. Harry and Naik, Aakash A. and Niblett, Samuel P. and Norwood, Sam Walton and O'Neill, Niamh and Ortner, Christoph and Persson, Kristin A. and Reuter, Karsten and Rosen, Andrew S. and Rosset, Louise A. M. and Schaaf, Lars L. and Schran, Christoph and Shi, Benjamin X. and Sivonxay, Eric and Stenczel, Tamás K. and Svahn, Viktor and Sutton, Christopher and Swinburne, Thomas D. and Tilly, Jules and van der Oord, Cas and Vargas, Santiago and Varga-Umbrich, Eszter and Vegge, Tejs and Vondrák, Martin and Wang, Yangshuai and Witt, William C. and Wolf, Thomas and Zills, Fabian and Csányi, Gábor},
	year = {2024},
}

@misc{Kingma2017,
	title = {Adam: {A} {Method} for {Stochastic} {Optimization}},
	shorttitle = {Adam},
	url = {http://arxiv.org/abs/1412.6980},
	doi = {10.48550/arXiv.1412.6980},
	urldate = {2026-06-01},
	publisher = {arXiv},
	author = {Kingma, Diederik P. and Ba, Jimmy},
	month = jan,
	year = {2017},
}

@article{Bannwarth2019,
	title = {{GFN2}-{xTB}—{An} {Accurate} and {Broadly} {Parametrized} {Self}-{Consistent} {Tight}-{Binding} {Quantum} {Chemical} {Method} with {Multipole} {Electrostatics} and {Density}-{Dependent} {Dispersion} {Contributions}},
	volume = {15},
	copyright = {http://pubs.acs.org/page/policy/authorchoice\_termsofuse.html},
	issn = {1549-9618, 1549-9626},
	url = {https://pubs.acs.org/doi/10.1021/acs.jctc.8b01176},
	doi = {10.1021/acs.jctc.8b01176},
	number = {3},
	urldate = {2026-06-01},
	journal = {Journal of Chemical Theory and Computation},
	author = {Bannwarth, Christoph and Ehlert, Sebastian and Grimme, Stefan},
	month = mar,
	year = {2019},
	pages = {1652--1671},
}

@article{Unke2021,
	title = {Machine {Learning} {Force} {Fields}},
	volume = {121},
	copyright = {https://creativecommons.org/licenses/by-nc-nd/4.0/},
	issn = {0009-2665, 1520-6890},
	url = {https://pubs.acs.org/doi/10.1021/acs.chemrev.0c01111},
	doi = {10.1021/acs.chemrev.0c01111},
	number = {16},
	urldate = {2026-06-05},
	journal = {Chemical Reviews},
	author = {Unke, Oliver T. and Chmiela, Stefan and Sauceda, Huziel E. and Gastegger, Michael and Poltavsky, Igor and Schütt, Kristof T. and Tkatchenko, Alexandre and Müller, Klaus-Robert},
	month = aug,
	year = {2021},
	pages = {10142--10186},
}

@article{Wild2025,
	title = {Automatic feature selection and weighting in molecular systems using {Differentiable} {Information} {Imbalance}},
	volume = {16},
	issn = {2041-1723},
	url = {https://www.nature.com/articles/s41467-024-55449-7},
	doi = {10.1038/s41467-024-55449-7},
	number = {1},
	urldate = {2026-06-05},
	journal = {Nature Communications},
	author = {Wild, Romina and Wodaczek, Felix and Del Tatto, Vittorio and Cheng, Bingqing and Laio, Alessandro},
	month = jan,
	year = {2025},
	pages = {270},
}

@article{Tipping,
	title = {Sparse {Bayesian} {Learning} and the {Relevance} {Vector} {Machine}},
	volume = {1},
	journal = {Journal of Machine Learning Research},
	author = {Tipping, Michael E},
	year = {2001},
	pages = {211--244},
}

@book{Rasmussen2008,
	address = {Cambridge, Mass.},
	edition = {3. print},
	series = {Adaptive computation and machine learning},
	title = {Gaussian processes for machine learning},
	isbn = {978-0-262-18253-9},
	publisher = {MIT Press},
	author = {Rasmussen, Carl Edward and Williams, Christopher K. I.},
	year = {2008},
}

@article{Veit2020,
	title = {Predicting molecular dipole moments by combining atomic partial charges and atomic dipoles},
	volume = {153},
	issn = {0021-9606, 1089-7690},
	url = {https://pubs.aip.org/jcp/article/153/2/024113/1061473/Predicting-molecular-dipole-moments-by-combining},
	doi = {10.1063/5.0009106},
	number = {2},
	urldate = {2026-06-07},
	journal = {The Journal of Chemical Physics},
	author = {Veit, Max and Wilkins, David M. and Yang, Yang and DiStasio, Robert A. and Ceriotti, Michele},
	month = jul,
	year = {2020},
	pages = {024113},
}

@article{Cho2026,
	title = {Benchmarking physics-inspired machine learning models for transition metal complexes with diverse charge and spin states},
	volume = {5},
	issn = {2635-098X},
	url = {https://xlink.rsc.org/?DOI=D5DD00571J},
	doi = {10.1039/D5DD00571J},
	number = {5},
	urldate = {2026-06-18},
	journal = {Digital Discovery},
	author = {Cho, Yuri and Briling, Ksenia R. and Calvino Alonso, Yannick and Laplaza, Ruben and Corminboeuf, Clemence},
	year = {2026},
	pages = {2103--2119},
}

@article{Batatia2025,
	title = {A foundation model for atomistic materials chemistry},
	volume = {163},
	issn = {0021-9606, 1089-7690},
	url = {https://pubs.aip.org/jcp/article/163/18/184110/3372267/A-foundation-model-for-atomistic-materials},
	doi = {10.1063/5.0297006},
	number = {18},
	urldate = {2026-06-18},
	journal = {The Journal of Chemical Physics},
	author = {Batatia, Ilyes and Benner, Philipp and Chiang, Yuan and Elena, Alin M. and Kovács, Dávid P. and Riebesell, Janosh and Advincula, Xavier R. and Asta, Mark and Avaylon, Matthew and Baldwin, William J. and Berger, Fabian and Bernstein, Noam and Bhowmik, Arghya and Bigi, Filippo and Blau, Samuel M. and Cărare, Vlad and Ceriotti, Michele and Chong, Sanggyu and Darby, James P. and De, Sandip and Della Pia, Flaviano and Deringer, Volker L. and Elijošius, Rokas and El-Machachi, Zakariya and Fako, Edvin and Falcioni, Fabio and Ferrari, Andrea C. and Gardner, John L. A. and Gawkowski, Mikołaj J. and Genreith-Schriever, Annalena and George, Janine and Goodall, Rhys E. A. and Grandel, Jonas and Grey, Clare P. and Grigorev, Petr and Han, Shuang and Handley, Will and Heenen, Hendrik H. and Hermansson, Kersti and Ho, Cheuk Hin and Hofmann, Stephan and Holm, Christian and Jaafar, Jad and Jakob, Konstantin S. and Jung, Hyunwook and Kapil, Venkat and Kaplan, Aaron D. and Karimitari, Nima and Kermode, James R. and Kourtis, Panagiotis and Kroupa, Namu and Kullgren, Jolla and Kuner, Matthew C. and Kuryla, Domantas and Liepuoniute, Guoda and Lin, Chen and Margraf, Johannes T. and Magdău, Ioan-Bogdan and Michaelides, Angelos and Moore, J. Harry and Naik, Aakash A. and Niblett, Samuel P. and Norwood, Sam Walton and O’Neill, Niamh and Ortner, Christoph and Persson, Kristin A. and Reuter, Karsten and Rosen, Andrew S. and Rosset, Louise A. M. and Schaaf, Lars L. and Schran, Christoph and Shi, Benjamin X. and Sivonxay, Eric and Stenczel, Tamás K. and Sutton, Christopher and Svahn, Viktor and Swinburne, Thomas D. and Tilly, Jules and Van Der Oord, Cas and Vargas, Santiago and Varga-Umbrich, Eszter and Vegge, Tejs and Vondrák, Martin and Wang, Yangshuai and Witt, William C. and Wolf, Thomas and Zills, Fabian and Csányi, Gábor},
	month = nov,
	year = {2025},
	pages = {184110},
}

\end{document}